\documentstyle[epsfig]{aipproc}

\begin{document}

\thispagestyle{empty}
\setcounter{page}{0}
\def\thefootnote{\fnsymbol{footnote}}

\begin{flushright}
IFT/00-36\\
hep-ph/0101111

\end{flushright}

\vspace{1cm}

\begin{center}

{\large\sc {\bf Phenomenology of the Chargino \\ and Neutralino Systems}}
\footnote{Talk given at the 5th International Linear 
Collider Workshop (LCWS 2000), Fermilab, Batavia, Illinois, Oct. 24-28, 2000}

\vspace{1cm}

{\sc J. Kalinowski
\footnote{
email: kalino@fuw.edu.pl
}%

\vspace*{1cm}
Instytut Fizyki Teoretycznej, Uniwersytet Warszawski\\ 
Ho\.za 69, 00681 Warsaw, Poland
}

\vspace*{1cm}

{\bf Abstract}
\end{center}
The chargino and neutralino pair
production processes at $e^+e^-$ collisions are explored to determine
the underlying SUSY parameters. 
The sum rules for the couplings are used to check the closure 
of two-chargino and 
four-neutralino systems in the minimal supersymmetric extension 
of the standard model. 

\def\thefootnote{\arabic{footnote}}
\setcounter{footnote}{0}

\newpage


\title{Phenomenology of the Chargino \\ and Neutralino Systems}

\author{Jan Kalinowski$^*$ }
\address{$^*$Instytut Fizyki Teoretycznej, Uniwersytet Warszawski\\ 
Ho\.za 69, 00681 Warsaw, Poland \\E-mail: kalino@fuw.edu.pl }

\maketitle

\begin{abstract}
The chargino and neutralino pair
production processes at $e^+e^-$ collisions are explored to determine
the underlying SUSY parameters. 
The sum rules for the couplings are used to check the closure 
of two-chargino and 
four-neutralino systems in the minimal supersymmetric extension 
of the standard model. 
\end{abstract}

Assuming that supersymmetry is realized in Nature \cite{hm}, its confusing
signals can be disentangled by combining LHC and LC data.
The unique environment of $e^+e^-$ collisions, with tunable incoming
energy and polarized beams in particular \cite{gmp}, will provide
means to ``measure'' the fundamental SUSY parameters independently of
theoretical assumptions. This will allow us to confront them with
relations following from e.g.\ grand unification theories \cite{bz}.

In many SUSY scenarios charginos ($\tilde{\chi}^\pm_i$) and
neutralinos ($\tilde{\chi}^0_i$) are among the lightest supersymmetric
particles. In the minimal model (MSSM) the chargino sector depends
only on the SU(2) gaugino mass $M_2$ (which can be chosen real and
positive), the higgsino mass parameter $\mu =|\mu| e^{i\phi_\mu}$, and
the ratio $\tan\beta(=v_2/v_1)$ of the vacuum expectation values of
the two neutral Higgs fields, while neutralino sector depends in
addition on the U(1) gaugino mass $M_1 =|M_1| e^{i\phi_1}$.  Therefore
the chargino and neutralino production processes $e^+ e^- \rightarrow
\tilde{\chi}^-_i \tilde{\chi}^+_j$, $ \tilde{\chi}^0_i
\tilde{\chi}^0_j$ may serve as a good starting point towards a
systematic and model--independent determination of the fundamental
SUSY parameters \cite{tf}-\cite{vw}.  This is conveniently done in two
steps first by going from the measurable quantities to physical masses
and mixing angles of charginos and neutralinos and then to the
fundamental MSSM parameters: $M_1$, $M_2$, $\mu$ and $\tan\beta$.  For
an alternative method of using a global fit, see e.g.\ \cite{bhlp}.

The chargino mass matrix is diagonalized by two different $2\times 2$
unitary mixing matrices $U_L$ and $U_R$ acting on left- and
right-chiral components of $\tilde{W}^-, \tilde{H}^-$. They are
parameterized by two mixing angles $\phi_L$ and $\phi_R$ and
CP-violating phases $\beta_{L,R}$.  The neutralino mass matrix is
diagonalized by a single $4\times 4$ unitary matrix $O_{ij}$ which
involves six mixing angles $\theta_i$ and ten CP-violating phases.
All physical observables can be written in terms of physical masses,
mixing angles and CP-phases, which in turn are uniquely determined by
the fundamental parameters {$M_1,M_2,\mu,\tan\beta$}.  Note however,
that not all arbitrarily chosen masses, mixing angles and CP-phases
are consistent with the MSSM \cite{km2}.

In $e^+e^-$ collisions the production of chargino pairs receives
contributions from $s$--channel $\gamma$ and $Z$ exchanges, and
$t$--channel $\tilde{\nu}_e$ exchange, while neutralino pairs from
$s$--channel $Z$ and $t$-- and $u$--channel selectron
$\tilde{e}_{L,R}$ exchanges. For both
$\tilde{\chi}^-_i\tilde{\chi}^+_j$ and
$\tilde{\chi}^0_i\tilde{\chi}^0_j$ processes the production amplitude
\begin{eqnarray}
{\cal A}[e^+e^-\rightarrow\tilde{\chi}_i\tilde{\chi}_j]
  =\frac{e^2}{s}Q^{ij}_{\alpha\beta}
   \left[\bar{v}(e^+)\gamma_\mu P_\alpha  u(e^-)\right]
   \left[\bar{u}(\tilde{\chi}_i) \gamma^\mu P_\beta 
               v(\tilde{\chi}_j) \right],
\label{eq:production amplitude}
\end{eqnarray}
after a Fierz transformation of the $t$-- and $u$--channel
contributions, is expressed in terms of four bilinear charges, defined
by the chiralities $\alpha,\beta=L,R$ of the lepton and
chargino/neutralino currents.  The corresponding bilinear charges for
the $\tilde{\chi}^-_i\tilde{\chi}^+_j$ production take the form
\begin{eqnarray}
&& Q^{ij}_{RL}=\delta_{ij} D_R + \alpha^L_{ij} F_R,
          \qquad
   Q^{ij}_{LL}=\delta_{ij} D_L  + \alpha^L_{ij} F_L, \nonumber \\
&& Q^{ij}_{RR}=\delta_{ij} D_R +\alpha^R_{ij} F_R,
	\qquad
   Q^{ij}_{LR}=\delta_{ij} (D_L+{\textstyle\frac{T_{\tilde{\nu}}}{4s^2_W}})
       + \alpha^R_{ij}(F_L-\, {\textstyle\frac{T_{\tilde{\nu}}}{4s^2_W}}),
\label{eq:[12]}
\end{eqnarray}
while in the case of $\tilde{\chi}^0_i\tilde{\chi}^0_j$ production they 
are
\begin{eqnarray}
&&Q^{ij}_{RL}=+{\textstyle \frac{D_Z}{c_W^2}}{\cal Z}_{ij}
              +T_{\tilde{e}_R}g_{Rij}, \qquad 
  Q^{ij}_{LL}=+{\textstyle \frac{D_Z}{s_W^2c_W^2}}
             (s_W^2 -{\textstyle\frac{1}{2}}){\cal Z}_{ij}
              -U_{\tilde{e}_L}g_{Lij},\nonumber\\ 
&&Q^{ij}_{RR}=-{\textstyle \frac{D_Z}{c_W^2}}{\cal Z}^*_{ij}
              -U_{\tilde{e}_R}g^*_{Rij},\qquad
  Q^{ij}_{LR}=-{\textstyle \frac{D_Z}{s_W^2c_W^2}}
          (s_W^2 -{\textstyle\frac{1}{2}}){\cal Z}^*_{ij}
              +T_{\tilde{e}_L}g^*_{Lij},
\end{eqnarray}
with $s$--, $t$--, and $u$--channel propagators 
$   D_L=1+{\textstyle \frac{D_Z}{s_W^2 c_W^2}}
      (s_W^2 -{\textstyle\frac{1}{2}})(s_W^2-{\textstyle\frac{3}{4}})$, 
$   F_L={\textstyle\frac{D_Z}{4s_W^2 c_W^2}}
      (s^2_W-{\textstyle\frac{1}{2}})$, 
$   D_R=1+\frac{D_Z}{c_W^2}(s_W^2-\frac{3}{4})$,
$   F_R=\frac{D_Z}{4c_W^2}$,
$D_Z=s/(s-m^2_Z+im_Z\Gamma_Z)$, 
$T_{a}=s/(t-m^2_{a})$, 
$U_{a}=s/(u-m^2_{a})$. The couplings
$\alpha^{L,R}_{ij}=\delta_{ij} (-1)^i\cos2\phi_{L,R}+(1-\delta_{ij})
\sin2\phi_{L,R}e^{-i\beta_{L,R}}$, $g_{Rij}={\textstyle
\frac{1}{c_W^2}}O_{i1}O^*_{j1}$, $g_{Lij}={\textstyle \frac{1}{4
s_W^2c_W^2}} (O_{i2}c_W+O_{i1}s_W)(O^*_{j2}c_W+O^*_{j1}s_W)$ and
${\cal Z}_{ij}={\textstyle \frac{1}{2}}
(O_{i3}O^*_{j3}-O_{i4}O^*_{j4})$ are written in terms of the mixing
angles $\phi_{L,R}$ and CP-phases $\beta_{L,R}$ (for charginos), and
the diagonalization matrix elements $O_{ij}$ (for neutralinos).  Note
that for the chargino case the $Q^{ij}_{\alpha\beta}$ are linear in
$s_{2L,R}\equiv \sin2\phi_{L,R}$ and $c_{2L,R}\equiv\cos 2\phi_{L,R}$,
and the phases $\beta_{L,R}$ will disappear in quantities not
sensitive to chargino helicities.  The $\tilde{\nu}_e$ exchange
contributes only to the $LR$ chargino amplitude. In contrast, in the
neutralino case the $Q^{ij}_{\alpha\beta}$ are quadratic in $O_{ij}$
and selectron exchanges contribute to all neutralino amplitudes.

We define the polar angle $\Theta$ and azimuthal angle $\Phi$ of the
produced $\tilde{\chi}_i$ in the reference frame given by the $e^-$
momentum direction as the $z$-axis and the $e^-$ transverse
polarization vector as the $x$-axis.  The polarized differential cross
section ${\rm d}\sigma^{ij}/{\rm d}\Omega[e^+e^- \rightarrow
\tilde{\chi}_i\tilde{\chi}_j]$, for the $e^-$ and $e^+$ polarization
vectors $P$=$(P_T,0,P_L)$ and $\bar{P}$=$(\bar{P}_T
\cos\eta,\bar{P}_T\sin\eta, -\bar{P}_L)$ respectively, is given by
\begin{eqnarray}
\frac{{\rm d}\sigma^{ij}}{{\rm d}\Omega}
  &=&\frac{\alpha^2}{16 s} \lambda^{1/2} [
     (1-P_L\bar{P}_L)\Sigma^{ij}_{\rm unp}+(P_L-\bar{P}_L)\Sigma^{ij}_{LL}
  +P_T\bar{P}_T\cos(2\Phi-\eta)\Sigma^{ij}_{TT}], 
\label{eq:diff} \\
\Sigma^{ij}_{\rm unp}&=& 4\,\{[1-(\mu^2_i - \mu^2_j)^2
                   +\lambda\cos^2\Theta]Q_1
                   +4\mu_i\mu_j Q_2+2\lambda^{1/2} Q_3\cos\Theta\},
                  \nonumber\\
\Sigma^{ij}_{LL}     &=& 4\,\{[1-(\mu^2_i - \mu^2_j)^2
                   +\lambda\cos^2\Theta]Q'_1
                   +4\mu_i\mu_j Q'_2+2\lambda^{1/2} Q'_3\cos\Theta\},
                  \nonumber\\
\Sigma^{ij}_{TT}     &=&-4\lambda \sin^2\Theta\,\, Q_5,
\end{eqnarray}
where $\lambda=(1-(\mu_i+\mu_j)^2)(1-(\mu_i-\mu_j)^2)$ is the
two--body phase space function, $\mu_i^2=m^2_{\tilde\chi_i}/s$.  The
quartic charges are
\begin{eqnarray}
&& Q_1 ={\textstyle \frac{1}{4}}[|Q_{RR}|^2+|Q_{LL}|^2
                       +|Q_{RL}|^2+|Q_{LR}|^2], 
\quad Q_2 ={\textstyle  \frac{1}{2}}{\rm Re}[Q_{RR}Q^*_{RL}
                       +Q_{LL}Q^*_{LR}], 
\nonumber\\
&& Q_3 ={\textstyle  \frac{1}{4}}[|Q_{RR}|^2+|Q_{LL}|^2
                       -|Q_{RL}|^2-|Q_{LR}|^2], \quad 
Q_5={\textstyle\frac{1}{2}}{\rm Re} [Q_{LR}Q^*_{RR}
                       +Q_{LL}Q^*_{RL}],
\nonumber \\
&& Q'_1={\textstyle \frac{1}{4}}[|Q_{RR}|^2+|Q_{RL}|^2
                        -|Q_{LR}|^2-|Q_{LL}|^2], \quad 
Q'_2={\textstyle \frac{1}{2}}{\rm Re}[Q_{RR}Q^*_{RL}
                        -Q_{LL}Q^*_{LR}], \nonumber \\
&& Q'_3={\textstyle \frac{1}{4}}[|Q_{RR}|^2+|Q_{LR}|^2
                        -|Q_{RL}|^2-|Q_{LL}|^2].
\end{eqnarray}
The charges $Q_1$ to $Q_5$ are manifestly P--even,  
while $Q'_1$, $Q'_2$ and $Q'_3$ are P--odd. 

Since charginos and heavier neutralinos decay mainly into the
invisible lightest neutralinos and SM fermion pairs, the production
angles $\Theta$ and $\Phi$ cannot be determined completely on an
event--by--event basis. Integrating over $\Theta$ and $\Phi$, the
dependence of the total production cross sections on beam polarization
can be exploited to extract information on the mixing states of
charginos/neutralinos and on masses of exchanged sneutrinos/selectrons
\cite{vw}.  For the purpose of determining fundamental parameters it
is enough to consider the following integrated polarization--dependent
cross sections as physical observables \cite{choi}:
\begin{eqnarray}
\sigma^{ij}_{R,L}&=&\int{\rm d}\Omega\,\;
\frac{{{\rm d}\sigma^{ij}}}{{{\rm d}\Omega}}
              [P_L= \pm P^{max}, \bar{P}_L=\mp\bar{P}^{max}], 
\label{eq:xsections}
\end{eqnarray}
where $P^{max}$ and $\bar{P}^{max}$ are the maximum
longitudinal polarizations of $e^-$ and $e^+$.


The two--state mixing of charginos and four--state mixing of
neutralinos lead to sum rules for the chargino and for the neutralino
couplings. They can be formulated in terms of the squares of the
bilinear charges.  This follows from the observation that the mixing
matrices are unitary.  For example, the following general sum rules
can be derived for the two--state chargino system at tree level:
\begin{eqnarray}
&&\Sigma_{i,j=1,2}|Q^{ij}_{\alpha\beta}|^2 =
2(|D_\alpha|^2+|F_\alpha|^2), \quad\qquad (\alpha\beta)=(LL,RL,RR) 
\label{csr1},\\
&&\Sigma_{i,j=1,2}|Q^{ij}_{LR}|^2=2
 (|D_L+{\textstyle\frac{T_{\tilde{\nu}}}{4s^2_W}}|^2+
       |F_L- {\textstyle\frac{T_{\tilde{\nu}}}{4s^2_W}}|^2).
\label{csr2}
\end{eqnarray}
The right--hand side of (\ref{csr1}) is independent of any
supersymmetric parameters, while (\ref{csr2}) involves the sneutrino
mass. Similarly, the corresponding sum rules for the neutralino case
can be derived. The validity of these sum rules is reflected in both
the quartic charges and the production cross sections although, due to
mass effects and the $t/u$--channel sfermion exchanges, they are more
involved. Only {\it asymptotically} at high energies the sum rules
(like in eqs.\ref{csr1},\ref{csr2}) can be transformed directly into
the sum rules for the total cross sections.  We find \cite{inp}
\begin{eqnarray}
\Sigma_{ij=1,2}\; \sigma( \tilde{\chi}^-_i
       \tilde{\chi}^+_j)=\frac{347\; 
          {\rm fb}^{-1}}{s/{\rm TeV}^2}, \qquad
\Sigma_{ij=1,..,4}\; \sigma ( \tilde{\chi}^0_i
       \tilde{\chi}^0_j)=\frac{323\;  
         {\rm fb}^{-1}}{s/{\rm TeV}^2 }.
\end{eqnarray}
Nevertheless, the fact that all the physical observables are functions
of elements of the mixing matrices enables us to relate the cross
sections with the mixing angles. This has been worked out explicitly
for the chargino case. The six polarized cross sections
$\sigma^{ij}_{L,R}$ can be written as linear combinations of six
formally independent variables $\{z_k\}=\{1,c_{2L},c_{2R},c^2_{2L},
c^2_{2R},c_{2L}c_{2R}\}$, with coefficients which are known functions
of chargino masses, sneutrino mass (taken from e.g. sneutrino pair
production) and other known parameters (the scattering energy, SM
couplings etc.).  Inverting, the variables $z_i$ are given in terms of
the observables $\sigma^{ij}_{L,R}$. Since $z_i$ are not independent,
viz.\ $z_4=z^2_2$, $z_5=z^2_3$, $z_6=z_2 z_3$, we obtain several
non--trivial relations among the observables in the chargino
sector. The failure of saturating any of these sum rules by the
measured cross sections would signal that the chargino two--state
$\{\tilde\chi_1^\pm$, $\tilde\chi_2^\pm\}$ system is not complete and
additional states mix in. If the sum rules are satisfied, the $z_2$
and $z_3$ are the required mixing parameters $\cos2\phi_L$ and
$\cos2\phi_R$.  Similar techniques can be worked out for the
neutralino system \cite{inp}.

It has recently been demonstrated 
\cite{choi} that if the collider energy is sufficient to produce the
light and heavy chargino states in pairs, the underlying fundamental
SUSY parameters, $M_2,|\mu|, \cos\phi_{\mu}$ and $\tan\beta$, can be
extracted {\it unambiguously} from chargino masses and production
cross sections with polarized electron beams. The chargino masses can
be determined very precisely from the sharp rise of the cross
sections.  Defining $\Delta =
m^2_{\tilde{\chi}^\pm_2}-m^2_{\tilde{\chi}^\pm_1}$, the fundamental
parameters are
\begin{eqnarray}
&&\tan\beta=[(4m^2_W
                     +\Delta \,
                      (\cos 2\phi_R-\cos 2\phi_L))/(4m^2_W
                     -\Delta \,
                      (\cos 2\phi_R-\cos 2\phi_L))]^{1/2},
\label{eq:tanb} \\
&& M_2={\textstyle \frac{1}{2}}
        [2(m^2_{\tilde{\chi}^\pm_2}+m^2_{\tilde{\chi}^\pm_1}-2m^2_W)
              -\Delta \, 
               (\cos 2\phi_R+\cos 2\phi_L)]^{1/2}, \nonumber\\
&& |\mu|={\textstyle \frac{1}{2}}
        [2(m^2_{\tilde{\chi}^\pm_2}+m^2_{\tilde{\chi}^\pm_1}-2m^2_W)
              +\Delta \, 
               (\cos 2\phi_R+\cos 2\phi_L)]^{1/2}, 
\label{eq:M2mu}\\
&&B= [ \Delta^2
                   -(M^2_2-\mu^2)^2-4m^2_W(M^2_2+\mu^2
                   +m^2_W\cos^2 2\beta)]/8 m_W^2M_2|\mu|\sin2\beta,
\label{eq:signmu}
\end{eqnarray}
where $B={\rm sign}(\mu)$ in CP-invariant and $B=\cos\phi_\mu$ in
CP-noninvariant theories. 

%
With the additional knowledge of the lightest neutralino mass, $M_1$
can be determined up to a two-fold ambiguity. This ambiguity can be
resolved e.g. by measuring neutralino production cross sections.

As an example we consider a CP-invariant MSUGRA point with
$\tan\beta=3$, $m_0=100$ GeV, $M_{1/2}=200$ GeV, sign$(\mu)>0$.
Assuming 1 $\sigma$ statistical errors for cross section measurements
at $\sqrt{s}=800$ GeV and 500 fb$^{-1}$ per left- and right-polarized
electron beams, and the mass of $\tilde{\chi}^\pm_1$ to be measured
with an accuracy of 50 MeV and masses of $\tilde{\chi}^\pm_2$ and
$\tilde{\chi}^0_1$ with 200 MeV, the following accuracies may be
expected: $M_2=152\pm1.8$ GeV, $\mu=316\pm0.9$ GeV,
$\tan\beta=3\pm0.7$ and $M_1=78\pm0.7$ GeV.  Slightly better errors
are obtained for a high $\tan\beta=30$ scenario, except for the
$\tan\beta$ parameter which this time is poorly determined as
$\tan\beta>20$.

In determining the above error of $\tan\beta$ the information that
$|\cos\phi_\mu|=1$ (i.e. no CP-violating phase $\phi_\mu$) has been
used \cite{ck}.  If the value of $\tan\beta$ can be measured more
precisely elsewhere (e.g. in the Higgs sector with $\sim 10$\% error),
then eq.(\ref{eq:signmu}) can be used to verify that
$\cos\phi_\mu=1\pm0.1$.


To summarize, the chargino and neutralino sectors can be analyzed
independently of each other. This is important since the structure of
the neutralino sector may potentially be very complex in theories
beyond the MSSM. The sum rules for the production cross sections
provide a consistency check of the underlying theoretical picture of
the chargino and neutralino systems and the experimental procedure in
extracting the fundamental parameters.  The discussion presented here
has been carried out at the tree level; the higher--order corrections
\cite{bh} (which include parameters from other sectors of the MSSM)
demand iterative expansions in global analyses at the very end.

\bigskip
\noindent{Acknowledgments:}
Invaluable discussions with my collaborators are
acknowledged. This work has been supported by the KBN Grant 2~P03B~060
18.

\end{document}